\documentclass[12pt,a4paper]{article}
\usepackage[T1]{fontenc}
\usepackage[utf8]{inputenc}
\usepackage{authblk}
\usepackage{amsfonts}
\usepackage{mathtools}
\usepackage{slashed}
\usepackage{amsmath,amssymb}
\usepackage{graphicx,xcolor}
\usepackage{cite}
\usepackage{hyperref}

\definecolor{refcol}{rgb}{0.9,0.1,0.1}
\hypersetup{colorlinks=true,linkcolor=blue,citecolor=refcol,urlcolor=cyan,linktocpage}

\usepackage[capitalize]{cleveref}
\usepackage{tikz}
\usetikzlibrary{shapes}
\usetikzlibrary{plotmarks}

\usepackage{setspace}

\usepackage[left=1.5cm,right=1.5cm,top=2cm,bottom=2cm]{geometry}
\allowdisplaybreaks
\newcommand{\be}{\begin{equation}}
\newcommand{\ee}{\end{equation}}
\newcommand{\bea}{\begin{eqnarray}}
\newcommand{\eea}{\end{eqnarray}}

\def\XXint#1#2#3{{\setbox0=\hbox{$#1{#2#3}{\int}$ }
\vcenter{\hbox{$#2#3$ }}\kern-.6\wd0}}

\newcommand{\Qp}{\mathbb{Q}_p}
\newcommand{\Zp}{\mathbb{Z}_p}

\begin{document}
\begin{titlepage}

\title{
{\Huge\bf Enhanced Symmetry}\\ 
{\Huge\bf of the $p$-adic Wavelets}
}

\bigskip\bigskip\bigskip\bigskip\bigskip

\author{{\bf Parikshit Dutta}${}^{1}$\thanks{{\tt parikshitdutta@yahoo.co.in}},
                    {\bf Debashis Ghoshal}${}^2$\thanks{{\tt dghoshal@mail.jnu.ac.in}}, \\  
                        {\bf and Arindam Lala}$^{2,3}$\thanks{{\tt arindam.physics1@gmail.com}}\\
\hfill\\
${}^1${\it Asutosh College, 92 Shyama Prasad Mukherjee Road,}\\
{\it Kolkata 700026, India}
\hfill\\              
${}^2${\it School of Physical Sciences, Jawaharlal Nehru University,}\\
{\it New Delhi 110067, India}
\hfill\\              
${}^3${\it Instituto de F\'{i}sica, Pontificia Universidad Cat\'{o}lica de Valpara\'{i}so,}\\
{\it Casilla 4059, Valpara\'{i}so, Chile}
     }

\date{%
%
\bigskip\bigskip
\begin{quote}
\centerline{{\bf Abstract}}
{\small
Wavelet analysis has been extended to the $p$-adic line $\Qp$. The $p$-adic wavelets are complex valued 
functions with compact support. As in the case of real wavelets, the construction of the basis functions is recursive, 
employing scaling and translation. Consequently, wavelets form a representation of the affine group generated by 
scaling and translation. In addition, $p$-adic wavelets are eigenfunctions of a pseudo-differential operator, as a 
result of which they turn out to have a larger symmetry group. The enhanced symmetry of the $p$-adic wavelets 
is demonstrated. 
}
\end{quote}
}

\bigskip

\end{titlepage}
\thispagestyle{empty}\maketitle\vfill \eject

\tableofcontents

\section{Introduction}\label{Intro} 

Wavelet analysis may be considered as one of the cornerstones of modern day multi-resolution analysis for linear
and non-linear processes. It is a useful method in the digital signal processing and pattern recognition. However, 
wavelet analysis also finds applications in pure mathematics\cite{W_math}, statistical data analysis\cite{W_sda}, 
quantum field theory\cite{W_qft}, nonlinear dynamics\cite{W_nld}, to name a few. It is extremely useful in cases 
where traditional Fourier analysis is not very efficient. This is because wavelets are localised in both time and 
frequency, a property that resembles real life signals. In this approach, functions are expanded in terms of a set 
of basis functions known as  wavelet functions, which are compactly supported in frequency and time domain. 
This allows for local processing of functions independently at different scales. 

The mathematical description of wavelet theory begins with the representation of a square integrable function in 
terms of a complete set of orthonormal basis (although in practical applications, it is often more efficient to use a 
nonorthogonal set of functions). The first basis of this type was introduced by Haar\cite{Haar:1910} for  
$L^{2} \left( \mathbb{R} \right)$ 
long before the notion of wavelet transform was developed. The Haar wavelets, as they are called now, consist of 
dyadic translation and dilatation of a basic compactly supported piecewise constant function, known as the (Haar) 
mother wavelet function. Morlet and Grossmass's pioneering work on wavelet analysis \cite{MorletGrossmann:1984} 
was based on seismic wave data analysed by Morlet and collaborators and used smooth basis functions with 
compact support. Later work by Mallat, Meyer, Daubechies \cite{Mallat:1987,Mallat:1988,Meyer:1992,Daubechies:1992} 
and others developed the modern theory of wavelet transform (see \cite{Debnath:1998} for a brief historical review). 

The basis functions in wavelet analysis are constructed by application of scaling and translation on a mother wavelet, 
as a result of which all the basis functions have the same shape. Let $\Psi_M(x)$ be the mother wavelet. 
Then the scaled and translated functions are
\begin{equation}
\Psi_{a,b}(x) = \frac{1}{\sqrt{a}}\,\Psi_M\left(\frac{x-b}{a}\right),
\qquad a\in{\mathbb{R}}^+\:\text{ and }\: b\in{\mathbb{R}} 
\label{ScaleTransl}
\end{equation}
The functions $\Psi_{a,b}(x)$ may be thought to have resulted from $\Psi_M(x)$ due to the action of the elements 
$g(a,b)$, which satisfy
\begin{equation}
g(a_1,b_1)\, g(a_2,b_2) = g(a_1 a_2, b_1 + a_1 b_2)
\label{gpmult}
\end{equation}
and thus form the affine group `$ax+b$', a semi-direct product of the group of scaling and translation. 
In this notation $\Psi_M(x)=\Psi_{1,0}(x)$, which for the Haar wavelet is 
\begin{equation}
\Psi_M^{\text{Haar}}(x) = \left\{
\begin{array}{cl}
+1 & \text{for }\: 0\le x < \frac{1}{2}\\ 
-1 & \text{for }\: \frac{1}{2}\le x < 1\\
0 & \text{otherwise}
\end{array}
\right. 
\label{MotherHaar}
\end{equation}
An orthonormal basis may be obtained by restricting the scaling parameter $a$ to, say, the dyadic numbers
$2^{n}$ ($n\in{\mathbb{Z}}$), and the translation parameter $b\in\mathbb{Z}$. These are 
the original Haar wavelets. One may also get an orthonormal set of basis wavelets in the interval $\left[0,1\right]$
by restricting $n$ to negative integers and translations appropriately. 
The bases have the structure of a binary rooted tree with the mother wavelet $\Psi_{1,0}$ at the base. In 
the first generation, we have $\Psi_{\frac{1}{2},0}$ and $\Psi_{\frac{1}{2},1}$. In the second, $\Psi_{\frac{1}{2^2},0}$ 
and $\Psi_{\frac{1}{2^2},1}$ are connected to $\Psi_{\frac{1}{2},0}$, and $\Psi_{\frac{1}{2^2},2}$ and 
$\Psi_{\frac{1}{2^2},3}$ are connected to $\Psi_{\frac{1}{2},1}$, and so on. 

A straightforward generalisation of the dyadic case would start from the mother wavelet that assumes the values 
$\{1,\omega,\omega^2,\cdots,\omega^{p-1}\}$ (where $\omega$ is a primitive $p$-th root of unity) in the $p$ 
segments $\left[\frac{\ell}{p},\frac{\ell+1}{p}\right]$ ($\ell=0,\cdots,p-1$) of the interval $[0,1]$. In order to get an 
orthonormal basis $\left\{ \Psi^{(\text{H})}_{\frac{1}{p^n},m}\right\}$, we now restrict the scaling 
parameter to $p^{n}$ ($n\in{\mathbb{Z}}$) and the translations $m\in\mathbb{Z}$.

\medskip

Closely related to the real numbers is the field of $p$-adic numbers $\Qp$ obtained by the Cauchy completion of 
the rationals by a notion of distance derived from the non-archimedean $p$-adic norm
\cite{Koblitz:1984,VVZ:1994,Roberts:2000}. This leads to a very different topology on $\Qp$. The
ultrametric field $\Qp$ and its extensions have found many applications in physics, from glassy systems (see 
Ref.\cite{RTV:1986} for a review) to quantum mechanics, quantum field theory (see the monograph
\cite{VVZ:1994} and references therein), string theory (e.g., Refs.
\cite{FO:1987,FW:1987,BFOW:1988,Zabrodin:1989,Ghoshal:2006} and citations therein) and more recently in a 
discrete approach to holographic duality 
\cite{Gubser:2016a,Heydeman:2016ldy,Gubser:2016b,Gubser:2017,Parikshit:2017}. 
Wavelet analysis on $\Qp$ was pioneered by Kozyrev \cite{Kozyrev:2001}, who proposed a basis of complex
valued compactly supported wavelet functions for $L^{2}(\Qp)$. Kozyrev's construction may be considered 
as a generalization of the Haar type wavelets to the $p$-adic case. This was further generalised in Refs.\
\cite{Albeverio:2006,Albeverio:2009,Albeverio:2011} (a recent review is Ref.\cite{Kozyrev:2014}).

Interestingly Kozyrev's wavelets are eigenfunctions\cite{Kozyrev:2001} of the (generalised) \emph{Vladimirov 
derivative} \cite{VVZ:1994}, a (family of) pseudo-differen\-tial operator(s) on $\Qp$. Together with the natural shift 
operators that relate wavelets at different scales, this suggests a possibility of extending (the scaling part of) the 
affine symmetry of the wavelets to a larger one. We recall that various extensions of (the scaling part of) the affine 
symmetry of wavelets on $\mathbb{R}$ have been suggested is Refs.\cite{Ludu:1998, Ludu:1998h}. These are all 
deformations of $\mathfrak{sl}$(2). Indeed, we also find symmetry algebras that is closely related to $\mathfrak{sl}(2)$. 
More specifically, we find an exact $\mathfrak{sl}(2)$ algebra, and a class of non-linear ones labelled by one complex
parameter. The former (undeformed) algebra is obtained for the limiting value zero of the deformation parameter.  
In fact, this is not surprising given the fact that in the context of $p$-adic adS/CFT, the conformal field theories (CFT) 
on $\Qp$, or its algebraic extensions, are indeed analogues of one-dimensional CFTs\cite{Parikshit:2017} on the real 
line. An SL$\left( 2, \mathbb{R} \right)$ symmetry is associated with the latter \cite{dAFF:1976} (see also 
\cite{Jackiw:2011}).  In this approach, $\Qp$ (or its algebraic extension) is the asymptotic boundary of the Bruhat-Tits 
tree, which plays the analogue of the anti-de Sitter bulk space. It is is fact possible to extend this to an infinite dimensional
symmetry algebra that is formally similar to the Witt algebra of diffeomorphisms of a circle. We will demonstrate this, but
will postpone its connection with the symmetries of $\mathbf{P}(\Qp)$, and the large symmetries of the Bruhat-Tits tree,
for the future. The $p$-adic wavelets and their enhanced symmetry algebra should be the appropriate language for this 
CFT, and more generally for quantum field theories on ultrametric spaces and those associated with hierarchical 
models\cite{LernerMissarovQFT}. It may help in our understanding of systems where ultrametricity appear naturally. 
Examples of such systems are abound in nature. 

\section{A few results from $p$-adic analysis}
Let us recapitulate a few facts about the field $\Qp$ and aspects of $p$-adic analysis. More details are available in, 
e.g., Refs.\cite{Koblitz:1984,VVZ:1994,Roberts:2000}. First, let us fix a prime $p$. The $p$-adic
norm of an integer $n$ is $\left| n \right|_p = p^{-\text{ord}_p(n)}$, where ord$_p$ is the highest power of $p$ that 
divides $n$. Clearly, ord$_p$ behaves like a logarithm since ord$_p(n_1n_2)$ = ord$_p(n_1)$ + ord$_p(n_2)$. 
Hence the $p$-adic norm $|\cdot|_p$ of a rational number $m/n$ is defined as $\left|\displaystyle{\frac{m}{n}}
\right|_p = p^{\text{ord}_p(n) - \text{ord}_p(m)}$. The field $\Qp$ is the Cauchy completion obtained by including 
all the (suitably defined equivalence classes of) limits of converging sequences of rational numbers with respect to 
the $p$-adic norm. This norm has the ultrametric property leading to a stronger than usual triangle inequality 
$\left| \xi-\xi' \right|_p \leq \text{max}\left( |\xi|_p, |\xi'|_p\right)$. 

An element $\xi\in\Qp$ admits a Laurent series expansion in $p$: 
\begin{equation}
\xi = p^{N}\left(\xi_0+\xi_1p+\xi_2p^2+\cdots\right) = p^{N}\sum_{n=0}^\infty\xi_n p^n \label{plaurent}
\end{equation}
where $N\in\mathbb{Z}$ and $\xi_n\in\{0,1,\cdots,p-1\}$, $\xi_0\ne 0$. (There is, however, nothing special about this 
choice; one may work with other representative elements for the coefficients $\xi_n$.)  The expansion in $p$ above 
is convergent in the $p$-adic norm. The subset $\Zp$ consisting of elements with $N\geq 0$ in \cref{plaurent} (i.e., 
elements with norm less than equal to 1) is a subring known as the $p$-adic integers $\Zp$.

The finite part of the series \cref{plaurent}, $\displaystyle\sum_{n=N}^{-1}\xi_{n-N}p^{n}$, consisting only of the negative
powers of $p$, is called the \emph{fractional part} $\{\xi\}_p$ of $\xi$. The rest, an infinite series in general, is the 
\emph{integer part} $[\xi]_p$.  Let us define the complex valued function $\chi:\Qp\to\mathbb{C}$, as
\begin{equation}\label{pexponential}
\chi^{(p)}(\xi) = \exp\left(2\pi i\xi\right) = \exp\left(2\pi i\{\xi\}_p\right) 
\end{equation}
As in the real case, the contribution to \cref{pexponential} from the integer part $[\xi]_p$ is trivial and only the fractional 
part matters. Since $\chi^{(p)}(\xi+\xi')=\chi^{(p)}(\xi)\chi^{(p)}(\xi')$, it is called an {\it additive character} of $\Qp$. The 
totally disconnected nature of the $p$-adic space means that the (complex valued) continuous functions on $\Qp$ are 
locally constant. One such function is the \emph{indicator function} 
\begin{equation}
\Omega^{(p)}(\xi-\xi') = \left\{ 
\begin{array}{ll}
1 & \text{for } \left|\xi-\xi'\right|_p \le 1\\
0 & \text{otherwise}
\end{array}
\right. \label{indicator}
\end{equation}
and similarly for other open sets.

Using the additive character, one can define the Fourier transform of a complex valued function
\begin{equation}\label{pFT}
\widetilde{f}(\omega) = \int d\xi \chi^{(p)}\left(-\omega \xi\right) f(\xi)
\end{equation}
where $d\xi$ is the translationally invariant Haar measure on $\Qp$ normalized as $\displaystyle{\int_{\Zp}} 
d\xi=1$. The inverse Fourier transformation is given by
\begin{equation}\label{pinvFT}
f(\xi) = \int d\omega\,\chi^{(p)}\left(\omega\xi\right)\widetilde{f}(\omega) 
\end{equation}
The proof, available in the references cited, will be omitted.
The Fourier transform of an indicator function is also an indicator function. In this sense, these are analogues of 
the Gaussian functions on the real line.

\bigskip

The totally disconnected toplogy of $\Qp$ does not not allow for the usual definition of a derivative. The generalised 
Vladimirov derivative, therefore, is defined as an integral kernel
\begin{equation}\label{VladD}
D^\alpha f(\xi) = \frac{1-p^\alpha}{1-p^{-1-\alpha}}\, \int d\xi'\, \frac{f(\xi')-f(\xi)}{|\xi'-\xi|_p^{1+\alpha}}
\end{equation}
This expression is in fact meaningful for any $\alpha\in\mathbb{C}$ \cite{Albeverio:2011}. Furthermore, $D^{\alpha_1} 
D^{\alpha_2} = D^{\alpha_2} D^{\alpha_1} = D^{\alpha_1+\alpha_2}$.

In the following, we shall need a map from $\Qp$ to $\mathbb{R}$, which was introduced in Ref.\cite{Monna:1952}.
Following Kozyrev \cite{Kozyrev:2001}, we modify the Monna map slightly to define
\begin{align}\label{monna1}
\begin{split}
\mu : \: \Qp &\rightarrow\mathbb{R}_+\\
\sum_{m=N}^\infty \xi_m p^m &\mapsto \sum_{m=N}^\infty \xi_m p^{-m-1} 
\end{split}
\end{align}
This map, although, continuous, is not one-to-one. However, the induced map $\mu : \Qp/\Zp \rightarrow
\mathbb{N}\cup\{ 0\}$, is one-to-one. 

\section{A brief review of Kozyrev wavelets on $\Qp$} 
The notion of Haar wavelets was generalised to the case of $\Qp$ by Kozyrev in Ref.\cite{Kozyrev:2001}.  The set
of functions
\begin{align}
\label{padic_wavelet}
\psi_{n,m,j}^{(p)} (\xi) = p^{-\frac{n}{2}} \chi^{(p)} \left( j p^{n-1} \xi \right) \Omega^{(p)} \left( | p^n \xi - m |_{p} \right), 
\quad \xi \in \Qp
\end{align}
where $n\in\mathbb{Z}$, $m \in \Qp/\Zp$ and $j \in \{1,2,3,\cdots,p-1\}$. These functions satisfy
\begin{eqnarray}
\int_{\Qp} \psi_{n,m,j}^{(p)}(\xi) \, d\xi = 0
\end{eqnarray}
therefore belong to the set of mean zero locally constant functions $\mathbb{D}_0(\Qp)$. The wavelets provide 
an orthonormal basis for $L^2(\Qp)$ 
\begin{equation}
\int_{\Qp} \psi_{n,m,j}^{(p)}(\xi)\, \psi_{n', m',j'}^{(p)}(\xi)\, d\xi = \delta_{n,n'}\delta_{m,m'}\delta_{j,j'}
\label{orthognality}
\end{equation}
These properties are exactly analogous to the Haar wavelets on $\mathbb{R}$. 

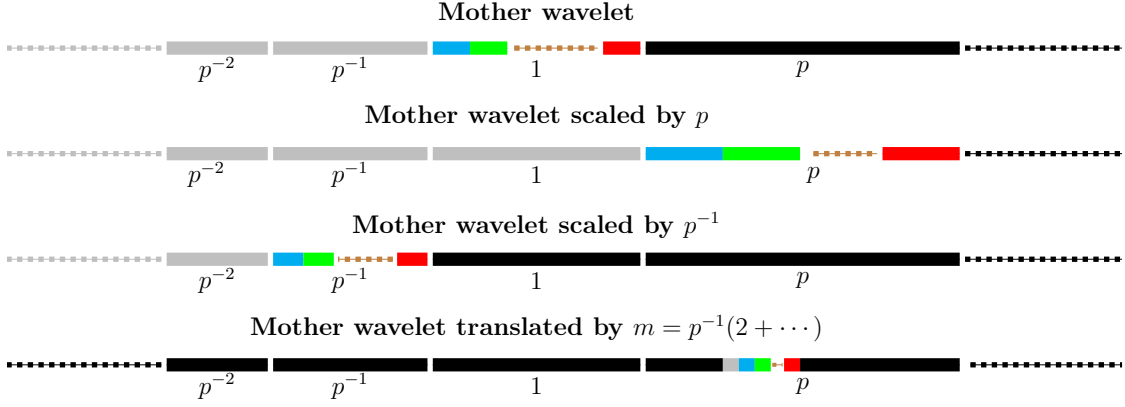
\begin{figure}[ht]
\centering
\begin{tikzpicture}[scale=0.70,every node/.style={scale=0.8}]
\filldraw[line width=5, cyan] (-1.95,2.5) -- (-1.25,2.5);
\filldraw[line width=5, green] (-1.25,2.5) -- (-0.55,2.5);
\filldraw[line width=2,dotted, brown] (-0.42,2.5) -- (1.15,2.5);
\filldraw[line width=5, red] (1.25,2.5) -- (1.95,2.5);
\filldraw[line width=5, black] (2.05,2.5) -- (7.95,2.5);
\filldraw[line width=2, dotted, black] (8.05,2.5) -- (10,2.5);
\filldraw[line width=2,dotted, black] (10.05,2.5) -- (11,2.5);
\filldraw[line width=5, lightgray] (-2.05,2.5) -- (-4.95,2.5);
\filldraw[line width=5, lightgray] (-5.05,2.5) -- (-6.95,2.5);
\filldraw[line width=2, dotted, lightgray] (-7.05,2.5) -- (-9.95,2.5);


\node at (0,2.1) {$1$};
\node at (5,2.1) {$p$};
\node at (-3.5,2.1) {$p^{-1}$};
\node at (-6,2.1) {$p^{-2}$};

\node at (0,3.2) {\bf Mother wavelet};

\filldraw[line width=5, lightgray] (-1.95,.5) -- (1.95,.5);
\filldraw[line width=5, cyan] (2.05,.5) -- (3.5,.5);
\filldraw[line width=5, green] (3.5,.5) -- (4.95,.5);
\filldraw[line width=2,dotted, brown] (5.2,.5) -- (6.40,.5);
\filldraw[line width=5, red] (6.5,.5) -- (7.95,.5);
\filldraw[line width=2, dotted, black] (8.05,.5) -- (11.0,.5);
\filldraw[line width=5, lightgray] (-2.05,.5) -- (-4.95,.5);
\filldraw[line width=5, lightgray] (-5.05,.5) -- (-6.95,.5);
\filldraw[line width=2, dotted, lightgray] (-7.05,.5) -- (-9.95,.5);


\node at (0,0.1) {$1$};
\node at (5.2,0.1) {$p$};
\node at (-3.5,0.1) {$p^{-1}$};
\node at (-6.2,0.1) {$p^{-2}$};

\node at (0,1.2) {\bf Mother wavelet scaled by $p$};

\filldraw[line width=5, black] (-1.95,-1.5) -- (1.95,-1.5);
\filldraw[line width=5, black] (2.05,-1.5) -- (7.95,-1.5);
\filldraw[line width=2, dotted, black] (8.05,-1.5) -- (11,-1.5);

\filldraw[line width=5, red] (-2.05,-1.5) -- (-2.62,-1.5);
\filldraw[line width=5, cyan] (-4.95,-1.5) -- (-4.38,-1.5);
\filldraw[line width=5, green] (-4.38,-1.5) -- (-3.81,-1.5);
\filldraw[line width=2,dotted, brown] (-2.70,-1.5) -- (-3.73,-1.5);

\filldraw[line width=5, lightgray] (-5.05,-1.5) -- (-6.95,-1.5);
\filldraw[line width=2, dotted, lightgray] (-7.05,-1.5) -- (-9.95,-1.5);

\node at (0,-1.9) {$1$};
\node at (5,-1.9) {$p$};
\node at (-3.5,-1.9) {$p^{-1}$};
\node at (-6,-1.9) {$p^{-2}$};

\node at (0,-0.85) {\bf Mother wavelet scaled by $p^{-1}$};

\filldraw[line width=5, black] (-1.95,-3.5) -- (1.95,-3.5);
\filldraw[line width=5, black] (-4.95,-3.5) -- (-2.05,-3.5);
\filldraw[line width=5, black] (2.05,-3.5) -- (3.5,-3.5);
\filldraw[line width=2, dotted, black] (8.15,-3.5) -- (11,-3.5);
\filldraw[line width=5, black] (4.95,-3.5) -- (7.95,-3.5);

\filldraw[line width=5, lightgray] (3.5,-3.5) -- (3.8,-3.5);
\filldraw[line width=5, cyan] (3.8,-3.5) -- (4.1,-3.5);
\filldraw[line width=5, green] (4.1,-3.5) -- (4.4,-3.5);
\filldraw[line width=1.5, dotted, brown] (4.43,-3.5) -- (4.62,-3.5);
\filldraw[line width=5, red] (4.65,-3.5) -- (4.95,-3.5);

\filldraw[line width=5, black] (-5.05,-3.5) -- (-6.95,-3.5);
\filldraw[line width=2, dotted, black] (-7.05,-3.5) -- (-9.95,-3.5);

\node at (0,-3.9) {$1$};
\node at (5,-3.9) {$p$};
\node at (-3.5,-3.9) {$p^{-1}$};
\node at (-6,-3.9) {$p^{-2}$};

\node at (0,-2.8) {\bf Mother wavelet translated by $m=p^{-1}(2+\cdots)$};
\end{tikzpicture}
\caption{{\small A schematic representation of the wavelets. The sets are ordered by the values $|\xi|_p = p^n$.
(Colour code as follows: grey = 1, black = 0, other colours correspond to primitive roots of unity.)}}
\label{fig:wavelets}
\end{figure}

The mother wavelet $\psi_{0,0,1}^{(p)}(\xi)$ is supported on $\Zp$. Its value in this subset is 1 everywhere except on 
$\mathbb{S}_1$, i.e., $p$-adic numbers with norm exactly equal to 1. The open set $\mathbb{S}_1$, is a union of
open sets labelled by the value of $\xi_0$ in \cref{plaurent}, where it assumes the values $\omega_p^{j\xi_0}$ 
($\omega_p$ is a primitive $p$-th root of unity). After scaling $\psi_{n,0,j}^{(p)}(\xi)$ is supported on $|\xi|_p\le p^n$, 
where it is $p^{-{n}/{2}}$ for all $|\xi|_p < p^n$, and takes the values $p^{-\frac{n}{2}}\omega_{p}^{j\xi_0}$ in 
the set $|\xi|_p = p^n$, depending on the value of $\xi_0$
\begin{equation}
\psi_{n,0,j}^{(p)}(\xi) = \left\{\begin{array}{ll}
p^{-n/2} & \mathrm{ for }\; |\xi|_p \le p^{n}\\
p^{-n/2}\omega_p^{j\xi_0} &  \mathrm{ for }\; |\xi|_p =p^n\\
0 & \mathrm{ for }\; |\xi|_p > p^n
\end{array}\right. \label{scaledKozy}   
\end{equation}
The effect of translation is more involved. Moreover, the actual transformation depends on the choice of representative. 
Since the parameter of translation $m$ takes values in $\Qp/\Zp$, let us take, e.g., $m = m_0 p^{-1} + \cdots$ (where 
the terms denoted by the dots, being equivalent to zero, will not matter). The wavelet 
$\psi_{0,m,j}^{(p)}(\xi)$ is supported on the subset labelled by $\xi_0=m_0$ of the open set $|\xi|_p=p$. This subset is 
a union of subsets labelled by the values on $\xi_1$, where the function assumes the values $\omega_{p^2}^{j m_0}
\omega_p^{j\xi_1}$:
\begin{equation}
\psi_{0,m\in p^{-1}\Zp,j}^{(p)}(\xi) = \left\{\begin{array}{ll}
\omega_{p^2}^{j\xi_0}\omega_p^{j\xi_1} &  \mathrm{ for }\; |\xi|_p =p\:\mathrm{ and }\;\xi_0 = m_0 \\
0 & \left\{\displaystyle{\mathrm{ for }\; |\xi|_p < p, |\xi|_p > p\:\mathrm{ and }\atop |\xi|_p=p, 
\:\mathrm{ but }\;\xi_0\ne m_0\hfill}\right.\end{array}\right. \label{transKozy}   
\end{equation}
Notice that the measure of the support of a wavelet does not change under translation. The form of the translated wavelet
with other parameters can similarly be worked out. For example, for $m\in p^{-2}\Zp$, the translated wavelet is supported 
in the subset of the set $|\xi|_p=p^2$ defined by $\xi_0 = m_0$ and $\xi_1=m_1$. A schematic graph of the wavelets is 
shown in the \cref{fig:wavelets}.


The close relation between the generalised Haar wavelets on $\mathbb{R}$ and the Kozyrev wavelets\footnote{The 
additional freedom in choosing the phase through $j$ in \cref{padic_wavelet} will not be important, therefore, we set 
$j=1$ in the rest of this section and the next one, to avoid clutter. Also the labels for the Haar wavelets are changed 
from the ones used in \cref{Intro} 
to conform to the labels for the Kozyrev wavelets.} on $\Qp$ can be established via the Monna map \cref{monna1}. 
As in \cite{Kozyrev:2001}), it lets us define
\begin{equation}
\mu\,:\,\psi_{n,m}^{(p)} \mapsto \Psi^{\text{(H)}}_{n, m}\label{KozyToHaar}
\end{equation}
where $\Psi^{\text{(H)}}_{n,m}$ are 
\begin{equation}
\Psi^{\text{(H)}}_{n, m} (x) = p^{\frac{n}{2}} \sum_{\ell=0}^{p-1} e^{\frac{2\pi i\ell}{p}}
\Omega_{\left[\frac{m+\ell}{p^n},\frac{m+\ell+1}{p^n}\right]}\label{genHaar}
\end{equation}
in which $\Omega_{[a,b]}$ is the \emph{indicator function} for the interval $[a,b]$, i.e., $\Omega_{[a,b]}=1$ for
$x\in[a,b]$ and vanishes otherwise. 

An important property of the Kozyrev functions is that they are eigenfunctions of the Vladimirov derivative 
\cite{Kozyrev:2001}
\begin{equation}
D^\alpha \psi_{n,m,j}^{(p)} (\xi) = p^{\alpha(1-n)} \psi_{n,m,j}^{(p)} (\xi) \label{VDonKozy}
\end{equation}
with eigenvalue $p^{\alpha(1-n)}$. This property of the Kozyrev wavelets suggests that they behave like homogeneous
functions on $\mathbb{R}$ with definite scaling property. In order to make it apparent let us rewrite the above as
\begin{equation}
D^\alpha \psi_{n,m}^{(p)} (a \xi + b) = |a|_p^{\alpha(1-n)} \psi_{n,m}^{(p)} (a\xi + b) \label{KozyScal}
\end{equation}
where we have emphasised the scaling behaviour that will play an important role in the next section.

\section{Enhanced symmetry of the Kozyrev wavelets}
By their construction, the Kozyrev wavelets on $\Qp$ , like the (generalised) Haar wavelets on $\mathbb{R}$ 
(or on $\left[0,1\right]\in\mathbb{R}$, are organised by the affine group `$ax+b$'. Specifically, one can define 
raising and lowering operators such that
\begin{align}
\label{raise_lower}
\begin{split}
a_\pm\Psi^{\text{(H)}}_{n,m}(x) &= \Psi^{\text{(H)}}_{n\pm 1,m}(x) ,\quad x\in\mathbb{R}\\
a^{(p)}_\pm\psi^{(p)}_{n,m}(\xi) &= \psi^{(p)}_{n\pm 1,m}(\xi), \quad \xi\in\Qp
\end{split}
\end{align}
As we have seen, in the $p$-adic case, there is an additional operator that acts naturally on the wavelets. It is the 
generalised Vladimirov derivative, the action of which is given in \cref{VDonKozy}. It is easy to see, by a straightforward
adaptation of the proof in \cite{Kozyrev:2001}, that in the limit\footnote{Our definition of $\log_p D = \ln D/\ln p$, where
$\ln D = \displaystyle\lim_{\alpha\to 0}\frac{1}{\alpha}\left(D^\alpha - 1\right)$ is standard.} $\alpha\to 0$, one gets
\begin{equation}
\left(\log_p\! D\right)  \psi_{n,m}^{(p)} (\xi) = (1-n) \psi_{n,m}^{(p)} (\xi) \label{logVD}
\end{equation}
Combining the actions of $\log_p D$ and the raising and lowering operators $a_\pm$,  let us define the shift 
operators $J^{(p)}_\pm = a_\pm\log_p\!D$. They act as
\begin{align}
\label{raise_lower}
\begin{split}
J^{(p)}_\pm\psi^{(p)}_{n,m}(\xi) &= (1-n)\, \psi^{(p)}_{n\pm 1,m}(\xi), \quad \xi\in\Qp
\end{split}
\end{align}
on the Kozyrev wavelets. 

The property \cref{KozyScal} is in fact analogous to the action of the scaling operator  $p^{-x\partial_x}$ on 
homogeneous functions on $\mathbb{R}$. Let us consider the elementary homogeneous function $x^s$, which, 
for simplicity we shall restrict to the interval $\left[0,1\right]\in\mathbb{R}$. We define the map $\rho$ that takes 
the set of Kozyrev wavelets $\psi_{n,m}^{(p)}(\xi)$ to the set of monomials $x^{n-1}$ defined on the interval $[0,1]$
\begin{equation}
\rho\left(\psi_{n,m}^{(p)}(\xi)\right) = x^{n-1}
\label{maprho} 
\end{equation}
where we recognise the exponent as $(n-1)=-2\log_p\left(\mathrm{max}|\psi_{n,m}^{(p)}|\right) -1$.

Applying this map on \cref{VDonKozy}, we see that
\begin{align}
\label{rhoD}
\begin{split}
\rho\left( D^{\alpha}\psi_{n,m}^{(p)}(\xi)\right) & = p^{\alpha(1-n)}\, \rho\left(\psi_{n,m}^{(p)}(\xi)\right)    \\
\rho\circ D^{\alpha}\circ \rho^{-1}\left( x^{n-1}\right) & = p^{\alpha(1-n)}\, x^{n-1} 
\end{split}
\end{align}
Therefore, we conclude that
\begin{equation}
\rho\circ D^{\alpha}\circ \rho^{-1} = p^{-\alpha x\partial_x} = 
\exp\left( -\alpha \ln p \frac{\partial}{\partial \ln x}\right) 
\end{equation}
In other words, the Vladimirov operator on $\Qp$ maps to the scaling operator on the monomials on 
$[0,1]\in\mathbb{R}$. 

Taking a cue from this, it is natural to make the following association
\begin{equation}
\rho\circ J^{(p)}_+\circ \rho^{-1} =  -x^2 \frac{\partial}{\partial x}\equiv L_+\quad\text{ and }\quad
\rho\circ J^{(p)}_-\circ \rho^{-1} =  -\frac{\partial}{\partial x}\equiv L_- 
\label{rhoJ}
\end{equation}
These operators, together with $L_0 = -x\partial_x$,  satisfy $\left[L_0,L_\pm\right] = \mp L_\pm$ and 
$\left[L_+,L_-\right] = 2L_0$ to generate the $\mathfrak{sl}(2,\mathbb{R})$ algebra. 

{}From their action on the Kozyrev wavelets, we find an exact $\mathfrak{sl}(2,\mathbb{R})$ algebra for the operators 
$J_\pm^{(p)}$ and $\log_p D$,
\begin{equation}
\label{pSL2}
\left[ J_+^{(p)}, J_-^{(p)} \right]  = 2\ln_p D,\qquad\qquad\quad 
\left[ \log_p D, J_\pm^{(p)} \right] = \mp J_\pm^{(p)}  
\end{equation}
This is consistent with what we would get for the operators $L_0$ and $L_\pm$ using the map $\rho$ on the operators 
on the $p$-adic side. 

Indeed, it appears that one can do better and get an infinite dimensional algebra. Consider the generators 
$\ell_n^{(p)} = (a_\pm)^n \log_p D$, where one takes $a_\pm$ depending on whether $n$ is positive or negative. Of
course $\ell_n$ with $n=0,\pm 1$ are the $\mathfrak{sl}(2,\mathbb{R})$ generators above. The commutation relations
\begin{equation}
\left[\ell^{(p)}_n,\ell^{(p)}_m\right] = (n-m)\ell^{(p)}_{m+n} \label{pWitt}
\end{equation}
follow in a straightforward manner. This is formally analogous to the Witt algebra associated with the diffeomorphism 
of the circle. While it is very suggestive, and it will be interesting to explore its relation to the symmetries of $\Qp$, we
shall leave a proper investigation of this for future.

Coming back to an arbitrary non-zero $\alpha\in\mathbb{C}$,
\begin{align}
\label{pCommutator}
\begin{split}
\left[ D^\alpha, J_\pm^{(p)} \right] = \left(1 - p^{\pm\alpha}\right)\, D^\alpha J_\pm^{(p)},
\quad &\qquad\quad \left[D^\alpha, \log_p D\right] = 0,\\
\left[ J_+^{(p)}, J_-^{(p)} \right]  = 2\ln_p D,\quad 
&\qquad\quad \left[ \log_p D, J_\pm^{(p)} \right] = \mp J_\pm^{(p)}  
\end{split}
\end{align}
These may also be rewritten compactly as
\begin{equation}\label{pCommutation_alt}
p^{ \pm \frac{ \alpha}{2}} D^{\alpha} J_{\pm}^{(p)} - p^{ \mp \frac{\alpha}{2}} J_{\pm}^{(p)} D^{\alpha} = 0,
\qquad p^{\frac{\alpha}{2}\left[J_+,J_-\right]} = D^\alpha\qquad (\alpha\ne 0)
\end{equation}
which involves a kind of a \emph{deformed} commutator $\left[ A, B \right]_{\text{def}}\equiv qAB - q^{-1}BA$, with 
$q=p^{\pm\alpha}$, as well as the exponential function of the ordinary commutator. These define a (non-linear) 
symmetry that is larger than the scaling part of the affine group. We would also like to point to Ref.\cite{Ludu:1998} 
where several possible non-linear extensions of the affine symmetry of the wavelets on the real line was proposed. 
Although the symmetry is enhanced in both cases, the algebraic structure of the enhanced symmetry is quite different. 
However, both are related to $\mathfrak{sl}(2)$. This is expected due to the exponential nature of scaling on the 
wavelet basis\cite{Ludu:1998}. 
 
\bigskip 
 
We shall now provide some evidence in favour of the $\mathfrak{sl}(2,\mathbb{R})$ algebra proposed above. Let us 
begin by expressing the monomial in the basis of generalised Haar wavelets on $[0,1]$ (this restriction is only to stay 
within the set of square integrable functions)
\begin{equation}
\label{basis}
x^{n-1} = \sum_{n'=-\infty}^{0}\sum_{m'=0}^{(|n'|-1)/p^{n'}} c^{(n-1)}_{n',m'} \Psi^{\text{(H)}}_{n',m'}(x) \: \in[0,1]
\end{equation}
where, the coefficients $c^{(n-1)}_{n',m'}$ are determined by using the orthonormality of the Haar wavelets:
\begin{align}
\begin{split}
\label{repen_a}
c^{(n-1)}_{n',m'} &= \int_{0}^{1}x^{n-1}\Psi^{\text{(H)}}_{n',m'}(x) dx\:
=\: p^{\frac{n'}{2}}\sum_{j=0}^{p-1}e^{\frac{2\pi ij}{p}}\int_{(m'+j)p^{-n'}}^{(m'+j+1)p^{-n'}}x^{n-1}dx  \\
&= \frac{p^{-(n-\frac{1}{2})n'}}{n}\sum_{j=0}^{p-1}e^{\frac{2\pi ij}{p}}\left[(m'+j+1)^{n}-(m'+j)^{n}\right]
\end{split}
\end{align}
Let us first consider $L_-$, which acts as a derivative, therefore, $L_-(x^{n-1}) = (n-1)x^{n-2}$. We 
want to show that $\displaystyle{\int_0^1 dx\,\Psi_{n',m'}(x)\frac{d}{dx}x^{n-1}} = (n-1) \displaystyle{\int_0^1 dx\,
\Psi_{n',m'}(x) x^{n-2}}$. The RHS essentially repeats the computation above:
\begin{align}
\label{intgen_monoder1}
\begin{split}
(n-1)&{}\int_0^1 x^{n-2} \Psi^{\text{(H)}}_{n',m'}(x) dx \:=\: (n-1)
\int_{\frac{m'}{p^{n'}}}^{\frac{m'+p}{p^{n'}}} x^{n-2} \Psi^{\text{(H)}}_{n',m'} dx \\
&= \left( \int_{\frac{m'}{p^{n'}}}^{\frac{m'+1}{p^{n'}}} 1\; +\; \int_{\frac{m'+1}{p^{n'}}}^{\frac{m'+2}{p^{n'}}} e^{\frac{2\pi i}{p}}
+ \cdots +\; \int_{\frac{m' + p -1}{p^{n'}}}^{\frac{m'+p}{p^{n'}}} e^{\frac{2\pi i(p-1)}{p}} \right) (n-1) x^{n-2} dx \\
&= p^{- \left( n - \frac{3}{2} \right)n'}  \Bigg( 
- {m'}^{n-1} + \left( m' + 1 \right)^{n-1} \left( 1 - e^{\frac{2\pi i}{p}}\right) + \cdots \\
&{} \qquad\qquad + \left(m' + p -1 \right)^{n-1} e^{\frac{2\pi i \left( p - 2 \right)}{p}} \left( 1 - e^{\frac{2\pi i}{p}} \right) 
+ e^{\frac{2\pi i (p-1)}{p}} \left( m' + p \right)^{n-1}   \Bigg)  
\end{split}
\end{align}
The LHS, on the other hand, is
\begin{align}
\label{intgen_monoder2}
\begin{split}
\int_0^1 \!\!\!dx \, \Psi^{\text{(H)}}_{n',m'} (x)  \frac{d}{dx} x^{n-1} 
&= \underbrace{\Big[  \Psi^{\text{(H)}}_{n',m'}(x) x^{n-1}  \Big]_0^1}_{\text{boundary terms}}\; 
-\; \int dx \, \frac{d\Psi^{\text{(H)}}_{n',m'}}{d x} x^{n-1}\\
&= - \int dx \, x^{n-1} p^{\frac{n'}{2}} \sum_{j = 0}^{p-1} e^{\frac{2\pi i j}{p}} \left[ \delta\left( x - \frac{j+m'}{p^{n'}} \right) - 
\delta\left( x - \frac{j+m'+1}{p^{n'}} \right)  \right]  \\
&= - p^{\frac{n'}{2}} \sum_{j = 0}^{p-1} e^{\frac{2\pi i j}{p}} \left[ \left( \frac{j+m'}{p^{n'}} \right)^{n-1} - 
\left( \frac{j+m'+1}{p^{n'}} \right)^{n-1}   \right]  
\end{split}
\end{align}
Modulo the boundary term\footnote{There is some ambiguity in the boundary terms. Instead of the way we have defined
it above one may also take $\displaystyle\int_0^1 dx \, \Psi^{\text{(H)}}_{n',m'} (x) \frac{d\varphi(x)}{dx} = 
\displaystyle\int_{\frac{m'}{p^{n'}}}^{\frac{m'+p}{p^{n'}}} dx \, \Psi^{\text{(H)}}_{n',m'} (x) \frac{d\varphi(x)}{dx}= \left[ 
\Psi^{\text{(H)}}_{n',m'}(x) \varphi(x) \right]_{\frac{m'}{p^{n'}}}^{\frac{m'+p}{p^{n'}}}  + \cdots$, which are many more
terms. Since the Haar wavelets are step functions, the boundary values are not well defined.} (which we have omitted 
in the last two lines), this is exactly the same as the RHS of \cref{intgen_monoder1}. The boundary term at $x=0$ 
vanishes, so it does at $x=1$, unless the support of the Haar wavelet extends to this boundary. In that case, it is not 
properly defined, and one may take it to be either zero or one, or anything in between. We choose to take it to be zero. 
This is our justification for ignoring the boundary term, however, there may well be a better argument. 

Let us now consider the effect of dilatation, under which $x\to\lambda x$ and we would like to choose $\lambda =
p^{-\alpha}$. There is, however, a problem, since the interval $[0,1] \rightarrow [0,\lambda]$ after scaling. Consequently, 
$p^{-\alpha x\partial_x} (x^{n-1}) = e^{-\alpha\ln p \frac{\partial}{\partial\ln x}} e^{(n-1)\ln x} = p^{\alpha(1-n)} x^{n-1}$, 
is now supported on $[0,p^{-\alpha}]$. 
\begin{align}
\label{commute_dila}
\begin{split}
\left(p^{-\alpha} x\right)^{n-1} & = \sum_{n',m'} c_{n',m'}^{(n-1)} \Psi^{\text{(H)}}_{n',m'}(p^{-\alpha} x) \\
& = \frac{p^{-\left(n-\frac{1}{2}\right)n'}}{n} \sum_{j=0}^{p-1}e^{\frac{2\pi ij}{p}} \left[(m'+j+1)^{n} -(m'+j)^{n}\right] 
p^{\frac{n'}{2}}\sum_{k=0}^{p-1}e^{\frac{2\pi i k}{p}} \Omega_{\left[\frac{k+m'}{p^{n'-\alpha}},
\frac{k+m'+1}{p^{n'-\alpha}}\right]}\\
& = p^{\alpha(1-n)}\, \frac{p^{-\left(n-\frac{1}{2}\right)(n'-\alpha)}}{n} \sum_{j=0}^{p-1} e^{\frac{2\pi ij}{p}}
\left[(m'+j+1)^{n} - (m'+j)^{n}\right] \\
&\qquad\qquad\qquad\times\: p^{\frac{n' - \alpha}{2}} \sum_{k=0}^{p-1} e^{\frac{2\pi i k}{p}}
\Omega_{\left[\frac{k+m'}{p^{n'-\alpha}}, \frac{k+m'+1}{p^{n'-\alpha}}\right]}  \\
& = p^{\alpha(1-n)}\, \sum_{n',m} c_{n'-\alpha, m'}^{(n-1)} \Psi^{\text{(H)}}_{n'-\alpha,m'}(x) 
\: = \: p^{\alpha(1-n)}x^{n-1}
\end{split}
\end{align}
The scaling transformation is meaningful for any $\alpha > 0$, however, the shift in subscripts (labels of the Haar
wavelets) seems strange. To understand it better, let us restrict $\alpha\in\mathbb{Z}$, and consider
for definiteness $\alpha=\ell$. Depending on whether $\ell\in\mathbb{Z}^\pm$, the unit interval is shrunk (respectively
expanded) to $[0,p^{-\ell}]$. As a result one needs fewer (respectively, more) basis elements. This is ensured by the
shift of the labels.  

One can consider the limit $\alpha\to 0$ as well. In this case, the interval remains the same, and we need to prove 
that $\displaystyle{\int_0^1 dx\,\Psi_{n',m'}(x)x\frac{d}{dx}x^{n-1}} = (n-1) \displaystyle{\int_0^1 dx\,\Psi_{n',m'}(x) x^{n-1}}$.
The RHS is again similar to \cref{intgen_monoder1}, and on the LHS we can perform an integration by parts following
the steps in \cref{intgen_monoder2}, to verify the equation above. The action of $L_+$ on the monomials can be checked
in the same way.

\section{Translations}
So far we have considered only the scaling part of the affine symmetry of the wavelets. Just like the real wavelets,  
the orbits of the $p$-adic wavelets under the action of translations and dilatations form the affine group 
\cite{Albeverio:2009,Kozyrev:2014}. The parameter of translation is restricted to $m\in\Qp/\Zp$. 
It is clear from the definition of the Vladimirov derivative \cref{VladD} and the form of the wavelet function 
\cref{padic_wavelet}, that the action of $D^\alpha$ commutes with translation. 

The action of $J_\pm^{(p)}$ scales the argument of the wavelet, and effect of translation is a semi-direct product with this 
scaling. Let us consider translation by $m=p^{-1}\left(m_0+\cdots\right)\in p^{-1}\Zp$ of the wavelet\footnote{In the 
following, we have reinstated the phase label $j$ once again.} $\psi_{n,0,j}^{(p)}(\xi)$ \cref{scaledKozy} obtained from 
the mother wavelet by a scaling by $p^n$.  We get 
\begin{equation}
\psi_{n,(m\in p^{-1}\Zp),j}^{(p)}(\xi) = \left\{\begin{array}{ll}
p^{-\frac{n}{2}}\omega_{p^2}^{j\xi_0}\omega_p^{j\xi_1} &  \mathrm{ for }\; |\xi|_p =p^{n+1}\:\mathrm{ and }\;\xi_0 = m_0 \\
0 & \left\{\displaystyle{\mathrm{ for }\; |\xi|_p \le p^n, |\xi|_p\ge p^{n+2}\:\mathrm{ and }\atop |\xi|_p=p^{n+1}, 
\:\mathrm{ but }\;\xi_0\ne m_0\hfill}\right.\end{array}\right. \label{TSKozy}   
\end{equation}
On the other hand, the translated wavelet $\psi_{0,m\in p^{-1}\Zp,j}^{(p)}(\xi)$ \cref{transKozy}, after a scaling 
is $\psi_{n,p^n m,j}^{(p)}(\xi)$. The form of this function depends on whether $n$ is negative or positive. For
$n<0$, it is
\begin{equation}
\psi_{n,p^{-|n|} m,j}^{(p)}(\xi - m) = \left\{\begin{array}{ll}
0 & \left\{\displaystyle{\text{ for } |\xi|_p \le 1, \: |\xi|_p \ge p^2 \text{ and } |\xi|_p=p\hfill\atop
 \text{ but } \xi_i \ne m_i  \text{ for some } i=0,1,\cdots,|n|}\right.\\
p^{\frac{|n|}{2}} &  |\xi|_p=p \text{ and } \xi_i = m_i \;\forall\, i=0,1,\cdots,|n|+1\\
p^{\frac{|n|}{2}}\omega_{p}^{j\left(\xi_{|n|+1} - m_{|n|+1}\right)}  &  |\xi|_p=p \text{ and } \xi_i = m_i \;
\forall\, i=0,1,\cdots,|n| \text{ but } \xi_{|n|+1} \ne m_{|n|+1} 
\end{array}\right.\label{STKozy1}   
\end{equation}
For $n>0$, the final functions are different for $n=1$ and $n\ge 2$. (This is because the translation parameter
$m$ is chosen to be in $p^{-1}\Zp$.) 
\begin{align}\label{STKozy2} 
\begin{split}
(n=1)\qquad \psi_{1,p m,j}^{(p)}(\xi - m) &= \left\{\begin{array}{ll}
p^{-\frac{1}{2}}\omega_p^{-jm_0} & \text{for }\; |\xi|_p \le 1\\
p^{-\frac{1}{2}}\omega_p^{j(\xi_0-m_0)} & \text{for }\; |\xi|_p =p \text{ and } \xi_0\ne m_0\\
p^{-\frac{1}{2}} & \text{for }\; |\xi|_p =p \text{ but } \xi_0 = m_0\\
0 & \text{for }\; |\xi|_p \ge p^2
\end{array}\right. \\  
(n\ge 2)\qquad \psi_{n,p^n m,j}^{(p)}(\xi - m) &= \left\{\begin{array}{ll}
p^{-\frac{n}{2}}\omega_p^{-jm_0} & \text{for }\; |\xi|_p < p^n\\
p^{-\frac{n}{2}}\omega_p^{j \xi_0} & \text{for }\; |\xi|_p =p^n\\ 
0 & \text{for }\; |\xi|_p > p^n
\end{array}\right. \\  
\end{split}
\end{align}
After a scaling by a positive power of $p$, the (scaled) translation parameter $p^n m$ is valued in $\Zp$, and 
hence is equivalent to no translation. The form of the resultant wavelets \cref{STKozy2} are, therefore, the same
(upto a phase) as the scaled wavelets \cref{scaledKozy}.

\section{Conclusions}
Kozyrev's wavelets on the $p$-adic line $\Qp$ are a natural generalisation of the Haar wavelets on the real line. 
By their recursive construction using scaling and translation, the wavelet basis form a representation of these 
transformations, specifically, the group generated by their semi-direct product. This is the affine group, or the `$ax+b$' 
group. However, the Kozyrev functions are also eigenfunctions of the generalised Vladimirov derivative, which is a 
pseudo-differential operator. We have argued that when combined with the action of the logarithm of the Vladimirov 
derivative $\log_p D$, which can be defined through a limiting procedure, the scaling part of the symmetry is enhanced 
to $\mathfrak{sl}(2,\mathbb{R})$. Moreover, there is a one parameter family of non-linear symmetries, labelled by 
the exponent $\alpha\in\mathbb{C}$ of the Vladimirov derivative $D$.

The behaviour of the Kozyrev wavelets as eigenfunctions of $D$, is similar to the homogeneous functions on 
$\mathbb{R}$ with definite scaling property. In fact, we have proposed a map from the wavelets to the monomials 
$x^{n-1}$, ($n\in\mathbb{N}$). This identification has been justified by expanding the monomials in terms of 
generalised Haar wavelets, however, it would be instructive to verify it with other wavelet basis. The scaling behaviour
suggests that these wavelets are the right basis to use in addressing questions in $p$-adic CFT. Indeed, as we have 
shown, it is quite straightforward to get an infinite dimensional algebra that is formally identical to the Witt algebra of 
the diffeomorphism of the circle. It will be a very interesting problem to explore its relation to the symmetries of $\Qp$.

The enhanced symmetries of the $p$-adic wavelets, and their natural connection with the generalised Haar wavelets 
on $\mathbb{R}$ suggests that the latter system should also have a larger symmetry. At a formal level, this follows
from combining with the map in \cref{KozyToHaar}. Specifically,
\begin{equation*}
\mu\circ a_\pm^{(p)}\circ \mu^{-1} = a_\pm
\end{equation*}
where, $a_\pm$ are raising and lowering operators in \cref{raise_lower}. Similarly,
\begin{equation*}
\left(\mu\circ D^\alpha \circ \mu^{-1}\right) \Psi^{(\mathrm{H})}_{n,m} = p^{\alpha(1-n)}  \Psi^{(\mathrm{H})}_{n,m} 
\end{equation*}
in which the $n$ in the exponent, as in \cref{maprho}, is $n = -2\log_p\left(\mathrm{max}|\psi_{n,m}^{(p)}|\right)$. 
The idea of enlargement of symmetry of real wavelets have been explored in \cite{Ludu:1998,Ludu:1998h}. On a 
wavelet system consisting of smooth functions, it should be possible to realise the symmetry operators as 
differential operators. A generalisation of these results to wavelets on multi-dimensional spaces $\Qp^n$ should
also be possible.

 
\bigskip

\noindent{\bf Acknowledgments:} This work is supported by the research grant no.~5304-2, {\em Symmetries and 
Dynamics: Worldsheet and Spacetime}, from the Indo-French Centre for Promotion of Advanced Research 
(IFCPAR/CEFIPRA). DG would like to acknowledge the hospitality of the Humboldt Stiftung during the 
\emph{Humboldt Colloquium} in Bengaluru, where these results were presented. We thank Arghya Chattopadhyay,
Suvankar Dutta, Ram Ramaswamy and Riddhi Shah for useful discussions.


\end{document}